\documentclass[12pt,a4paper]{article}

\usepackage{amsmath}
\usepackage{amssymb}
\usepackage{epsfig}
\usepackage{graphicx}
\usepackage{cite}
\usepackage{url}

\begin{document}


\title{\Large\textbf{Testing the Stability of the Solar Neutrino LMA Solution with a Bayesian Analysis}}
\author{
{\large B.L. Chen$^{1}$,
  H.L. Ge$^{1}$, C. Giunti$^{2}$, Q.Y. Liu$^{1}$}
\bigskip
\\
{ $^{1}$ {\small \it Department of Modern Physics, University of
Science
    and}
}
\\
{ {\small \it  Technology  of China, Hefei, Anhui 230026, China.}
}
\\
{ $^2$ {\small \it INFN, Sezione di Torino, and Dipartimento di Fisica Teorica,} } \\
{  {\small \it  Universit\`a di Torino, Via P. Giuria 1, I--10125
Torino, Italy}}
\\
{ {\small \it }}
 }

\date{}
\maketitle\vskip 12mm

\begin{abstract}
We analyze with the Bayesian method
the solar and KamLAND neutrino data
in terms of neutrino oscillations.
We show that Bayesian credible regions
with a flat prior in the $\tan^2\theta_{12}$--$\Delta{m}^2_{21}$ plane
strongly support the LMA solution,
in agreement with the usual chi-square analysis.
Other reasonable priors are considered in order to test
the stability of the LMA solution.
We show that priors
which favor small or large values of the mixing angle
lead to minor changes of the allowed LMA region,
affecting mainly its large-$\tan^2\theta_{12}$ part.

\end{abstract}

\vspace*{.5cm}


\newpage

\renewcommand{\thefootnote}{\arabic{footnote}}
\setcounter{footnote}{0}

\section{Introduction}

A fraction of the electron neutrinos produced by thermonuclear
reactions in the core of the Sun disappear during their travel to
the Earth. This deficit is the famous ``Solar Neutrino Problem'',
discovered almost forty year ago. There have been many attempts to
solve this puzzle during the years since its discovery. Some of
them were based on modifications of the solar model in order to
have a lower neutrino production, although there was a conflict
with the energy spectrum provided by the four first-generation
experiments Homestake \cite{CHLOR}, Kamiokande \cite{Kamiokande},
GALLEX/GNO \cite{GALLEX} and SAGE \cite{SAGE}. Recent experiments
(Super-Kamiokande \cite{SK} and SNO
\cite{sno1,sno2,saltprl,saltprc}) have improved dramatically the
precision of solar neutrino data. The Sudbury Neutrino Observatory
(SNO) experiment \cite{sno1,sno2,saltprl,saltprc}, in which the
fluxes of solar $\nu_{e}$, $\nu_{e}+\nu_{\mu}+\nu_{\tau}$ and
$\nu_{e}+0.15(\nu_{\mu}+\nu_{\tau})$ on the Earth have been
measured through charged-current (CC), neutral-current (NC) and
elastic scattering (ES) interactions, has shown that the solar
neutrino deficit is due to $\nu_{e}\rightarrow \nu_{\mu,\tau}$
transitions. The fit of all solar neutrino data favor the
so-called Large Mixing Angle (LMA) solution, in which the MSW
effect \cite{MSW} is operative inside the Sun. The LMA solution
has been confirmed by the reactor experiment KamLAND \cite{kldet},
which observed $ \bar\nu_{e} \rightarrow \bar\nu_{x} $
oscillations of reactor antineutrinos.

Working in the framework of three-neutrino mixing with a small $\theta_{13}$
(see Ref.~\cite{Bilenky:1998dt}),
we analyzed the solar neutrino data with the effective mixing
\begin{eqnarray}
\null & \null & \null \nu_e = \cos\theta_{12} \, \nu_1 + \sin\theta_{12} \,
\nu_2 \,, \nonumber
\\
\null & \null & \null \nu_a = -\sin\theta_{12} \, \nu_1 + \cos\theta_{12} \,
\nu_2 \,, \label{mixing}
\end{eqnarray}
where $\theta_{12}$  is the mixing angle and $\nu_1$, $\nu_2$ are the
massive neutrinos with masses $m_1$, $m_2$.
The flavor neutrino $\nu_a$ is a linear combination of
$\nu_{\mu}$ and $\nu_{\tau}$,
which are indistinguishable in solar neutrino experiments
and in the KamLAND experiment.
In fact, it is easy to show that
$
\nu_a
\simeq
\cos\theta_{23} \, \nu_{\mu} - \sin\theta_{23} \, \nu_{\tau}
$
(see Ref.~\cite{Giunti-Salerno-06-04}),
with
$ \theta_{23} \simeq \pi/4 $
from the Super-Kamiokande atmospheric neutrino data
\cite{hep-ex/0604011}.

In the framework of neutrino oscillations, a standard least-squares
analysis method is usually adopted to analyze the data
in the context of Frequentist Statistics
\cite{LMA,Global2}.
These analyses have shown that the solar neutrino problem
is solved by the LMA solution.
However, there is a different statistical approach, the Bayesian Probability Theory.
The Bayesian analysis of solar neutrino data has several
advantages over a Frequentist one
(see the discussions in Refs.~\cite{bay2,James:2002iq,Giunti:2002ir}).
Bayesian analyses of early solar neutrino data have been presented in
Refs.~\cite{bay2,bay1,Creminelli:2001ij,model}.

In this paper, we analyze the updated solar neutrino data and the
last KamLAND data. The solar and KamLAND neutrino experiments are
briefly described in section~\ref{sec:experiments}.
Section~\ref{sec:numerical} gives a short description of our
implementation of the standard least-squares method. In
section~\ref{sec:bayesian} we present our results for the Bayesian
allowed regions corresponding to a flat prior in the plane of the
oscillation parameters $\tan^2\theta_{12}$ and $\Delta{m}^2_{21}
\equiv m_2^2 - m_1^2$. In section~\ref{sec:work} we investigate
the stability of the Bayesian allowed regions for other reasonable
choices of the prior. Conclusions are presented in
section~\ref{sec:conclusions}.

\section{Solar neutrino experiments}
\label{sec:experiments}

The first experiment to observe solar neutrinos,
was designed and
started by R. Davis and his collaborators in the late 1960s.
This is the famous radiochemical chlorine experiment located in the
Homestake Gold mine in South Dakota \cite{CHLOR}. It is made of 615 ton of $\rm
C_2Cl_4$. The weak process used for the detection is
\begin{eqnarray}
\nu_e + {}^{37}\text{Cl} & \rightarrow &
        {}^{37}\text{Ar}+ e^-
\ ,
\end{eqnarray}
which has an energy threshold of 0.814 MeV.
Thus,
the event rate measured in the Homestake experiment is due mainly
to $^{8}\text{B}$ solar neutrinos,
with small contributions from $^{7}\text{Be}$ and CNO neutrinos.
Other radiochemical experiments are the gallium experiments
GALLEX/GNO \cite{GALLEX},
located in the Laboratori Nazionali del Gran Sasso in Italy,
and
SAGE \cite{SAGE}, located in an
underground laboratory at Baksan.
The reaction for the detection is
\begin{eqnarray}
\nu_e + {}^{71}\text{Ga} & \rightarrow &
        {}^{71}\text{Ge} + e^-
\ ,
\end{eqnarray}
which has an energy threshold of 0.233 MeV.
Due to such a low energy threshold, the gallium experiments can detect the
most abundant neutrinos from the initial $pp$ reaction in the Sun.
The calculated flux in this part of the neutrino spectrum is much
less sensitive to the input parameters of the SSM, and thus is
more reliable, than that of the $^{8}\text{B}$ neutrinos.

Kamiokande \cite{Kamiokande} and Super-Kamiokande \cite{SK}
are water Cherenkov detectors
located in the Kamioka mine in Japan.
In these
experiments solar $\nu_e$'s are detected through the elastic scattering process
\begin{eqnarray}
\nu_e + e^- & \rightarrow & \nu_e + e^- \ .
\end{eqnarray}
We use the high-statistics data of the
Super-Kamiokande experiment,
which is a 50 kton detector sensitive to
$^8\text{B}$ solar neutrinos
with an energy threshold of about 5 MeV.

The SNO experiment \cite{sno1,sno2,saltprl,saltprc} is located in
Sudbury, Ontario, Canada. It is a heavy-water Cherenkov
detector which can make simultaneous measurements of
the $^8\text{B}$ solar $\nu_e$ flux
and
the $\nu_{\mu,\tau}$ flux
produced by neutrino oscillations
through the charged-current
and neutral-current interactions on deuterons
and the elastic scattering on electrons
\begin{displaymath}
\begin{array}{l l l l}
\nu_e + d          & \rightarrow &
        p~ \!+ p~+ e^- & \qquad\text{(CC)}, \\
\nu_x + d          & \rightarrow &
        p + n + \nu_x        & \qquad\text{(NC)}, \\
\nu_x + e^-\!\!\!\!& \rightarrow &
        \nu_x + e^-                 & \qquad\text{(ES).}
\end{array}
\end{displaymath}
The first phase of the SNO experiment was carried out from Nov. 2,
1999 to Jan. 15, 2001 \cite{saltprl} using pure $\rm D_2O$.
The data collected during this phase
allowed the SNO collaboration to publish
in 2002 \cite{sno2}
a model-independent evidence of solar
$ \nu_e \to \nu_{\mu,\tau} $ transitions.
In the second phase, from July 26, 2001 to Aug. 28,
2003 \cite{saltprc}, 2 ton of NaCl were added to the $\rm D_2O$
target to enhance the detection efficiency of the NC channel.
The SNO salt-phase data confirmed the results of the $\rm D_2O$ phase,
with more precision.

Different from the solar neutrino experiments,
KamLAND \cite{kldet}
is an experiment which detects the
antineutrinos produced by the decay of heavy nuclei
in commercial Japanese and Korean nuclear reactors. It is a
1000 ton liquid scintillator detector located in the Kamioka mine
in Japan. KamLAND detects neutrinos though the inverse beta decay
process
\begin{eqnarray}
\bar\nu_e + p       & \rightarrow &
        \ e^+ + n \ ,
\end{eqnarray}
which has a 1.8 MeV energy threshold. KamLAND is the first
experiment which observd a disappearance of reactor $\bar\nu_e$'s.
The results of the KamLAND experiment provided a definitive proof in favor of
the LMA solution of the solar neutrino
problem \cite{Hol,LMA_Sun,Bahcall}.
In our analysis we use the
766 ton-years data \cite{kldet}.


\section{Standard least-squares method}
\label{sec:numerical}

Following the tradition \cite{LMA,Global2}, we first
analyzed the data with a standard least-squares analysis (often
called ``$\chi^2$ analysis'').
In our analysis, the least-squares function for each type of
solar neutrino experiment is
\begin{equation}\label{e1}
X^2 = \sum_{ij} \, (R_{\,i}^{\, exp}-R_{\,i}^{\, th}) \,
\sigma^{-2}_{\,ij} \, (R_{\,j}^{\, exp}-R_{\,j}^{\, th})\, ,
\end{equation}
where $R_{\,i}^{\, exp}$ and $R_{\,i}^{\, th}$ are, respectively, the
experimental values and theoretical predictions of the observables.
The theoretically calculated rates $R_{\,i}^{\, th}$
depend on the oscillation parameters $\Delta{m}^2_{21}$
and $\tan^2\theta_{12}$.
In Eq.~(\ref{e1}),
$\sigma^{\,-2}$ is the inverse of the covariance
error matrix built by adding in quadrature the statistical and systematic errors,
considering mutual correlations.
We have used the data of the Homestake, GALLEX/GNO, SAGE, Super-Kamiokande and SNO
solar neutrino experiments.
For the global analysis of solar neutrino data, we define the solar neutrino
least-squares function as
\begin{equation}\label{e2}
X^2_{\mathrm{S}} = X^2_{\mathrm{Cl, Ga}} + X^2_{\mathrm{SK}} + X^2_{\mathrm{SNO}}
\ ,
\end{equation}
where we have separated the least-squares
function of the radiochemical experiments,
in which only the total rate is measured,
and those of the Super-Kamiokande and SNO experiments,
which measured also the energy spectra in the ES and CC reactions.
The initial $^8\text{B}$ solar neutrino flux is considered
as a free parameter to be determined by the fit,
mainly through the SNO NC data.
For the other solar neutrino fluxes, we assume the
BP04 Standard Solar Model \cite{BP04}
and treat the uncertainties as described in Refs.~\cite{Fogli:1994nn,hep-ph/9912231,hep-ph/0006026,bay2}.

In the KamLAND experiment, the survival probability of the reactor
electron antineutrinos can be written as
\begin{eqnarray}
P({\overline \nu}_e \rightarrow {\overline \nu}_e) &= &
 1 -
\sin^22\theta_{12}\sin^2\left(\frac{1.27\Delta m^2 L}{E_{\nu}}\right), \label{prob}
\end{eqnarray}
where $L$ is the reactor-detector distance in meters,
$\Delta m^2$ is expressed in eV$^2$ and $E_{\nu}$ in MeV.
The results of the KamLAND
experiment are divided into 13 energy bins above the threshold.
In the analysis of KamLAND data, we
use the least-squares function \cite{fs}
\begin{equation}
\label{son}
X^2_{\mathrm{K}}=2 \sum_i \left[ (\eta N_i^{\,
th}- N_i^{\, exp})+ N_i^{\, exp} \ln\left(\frac{N_i^{\, exp}}{\eta
N_i^{\, th} }\right)\right] + \frac{(\eta - 1)^2}{\sigma_{\,sys}^2}
\ ,
\end{equation}
where
$N_i^{\, exp}$ and $N_i^{\, th}$
are, respectively, the measured and calculated event numbers in each energy bin.
The calculated event number $N_i^{\, th}$
depends on the oscillation parameters $\Delta{m}^2_{21}$
and $\tan^2\theta_{12}$.
In Eq.~(\ref{son}),
$\sigma_{\,sys}$ is the systematic uncertainty and $\eta$ is a free parameter
to be determined by the minimization of the least-squares function.

In our global analysis of the solar and KamLAND neutrino data, the least-squares function is
\begin{equation}\label{e3}
X^2 = X^2_{\mathrm{S}} + X^2_{\mathrm{K}}
\ .
\end{equation}


\section{Bayesian analysis}
\label{sec:bayesian}

In the Bayesian approach
it is assumed that there is some prior knowledge of the values of the
parameters to be determined by the data analysis.
This prior knowledge must be quantified by a
function which is called ``prior probability distribution function''
(see Refs.~\cite{bay2,bay1,Creminelli:2001ij,model}).
The Bayes' Theorem allows the calculation of the ``posterior probability distribution
function'',
which quantifies the knowledge of the values of the
parameters provided by the data viewed
in the light of the prior knowledge.

In the case of the effective two-neutrino mixing in Eq.~(\ref{mixing}),
the parameters to be determined with the statistical analysis of the
solar and KamLAND neutrino data are
$\tan^2\!\theta_{12}$ and $\Delta{m}^2_{21}$.

The Bayes' Theorem says that the posterior probability distribution
function of $\tan^2\!\theta_{12}$ and $\Delta{m}^2_{21}$ is given by
\begin{equation}
p(\tan^2\!\theta_{12},\Delta{m}^2_{21}|\mathrm{D},\mathcal{I}) = \frac{
p(\mathrm{D}|\tan^2\!\theta_{12},\Delta{m}^2_{21},\mathcal{I})
\,
p(\tan^2\!\theta_{12},\Delta{m}^2_{21}|\mathcal{I})
}{
p(\mathrm{D}|\mathcal{I}) } \,, \label{Bayes1}
\end{equation}
where $p(\mathrm{D}|\tan^2\!\theta_{12},\Delta{m}^2_{21},\mathcal{I})$ is the
likelihood function and $p(\tan^2\!\theta_{12},\Delta{m}^2_{21}|\mathcal{I})$
is the prior probability function.
$\mathrm{D}$ represents the data and $\mathcal{I}$
represents all the prior general knowledge and
assumptions on solar and neutrino physics
(in the Bayesian Probability Theory all probabilities are conditional).
The function $p(\mathrm{D}|\mathcal{I})$
is the global likelihood, which acts as a normalization constant
through the constraint
$
\int
\mathrm{d}\!\tan^2\!\theta_{12} \,
\mathrm{d}\Delta{m}^2_{21} \,
p(\tan^2\!\theta_{12},\Delta{m}^2_{21}|\mathrm{D},\mathcal{I})
= 1
$,
leading to
\begin{equation}
p(\mathrm{D}|\mathcal{I})
=
\int
\mathrm{d}\!\tan^2\!\theta_{12} \,
\mathrm{d}\Delta{m}^2_{21} \,
p(\mathrm{D}|\tan^2\!\theta_{12},\Delta{m}^2_{21},\mathcal{I})
\,
p(\tan^2\!\theta_{12},\Delta{m}^2_{21}|\mathcal{I})
\,.
\label{normalization}
\end{equation}

Assuming that the statistical and systematic errors have normal
distributions, the likelihood function is given by
\begin{equation}
p(\mathrm{D}|\tan^2\!\theta_{12},\Delta{m}^2_{21},\mathcal{I}) = \frac{
e^{-X^2_{\mathrm{S}}/2} }{
(2\pi)^{N_{\mathrm{S}}/2}\sqrt{|V_{\mathrm{S}}|} } \,  \frac{
e^{-X^2_{\mathrm{K}}/2} }{
(2\pi)^{N_{\mathrm{K}}/2}\sqrt{|V_{\mathrm{K}}|} } \,.
\label{sampling}
\end{equation}
Here $N_{\mathrm{S}}=41$ is the number of solar data points
and
$N_{\mathrm{K}}=13$ is the number of KamLAND data points.
$X^2_{\mathrm{S}}$ is the solar least-squares function in Eq.~(\ref{e1}) and
$V_{\mathrm{S}}$ is the corresponding covariance matrix.
$X^2_{\mathrm{K}}$ is the KamLAND least-squares function in Eq.~(\ref{son}) and
$V_{\mathrm{K}}$ is the corresponding covariance matrix.

\begin{figure}[t!]
\begin{center}
\begin{tabular}{cc}
\mbox{\epsfig{figure=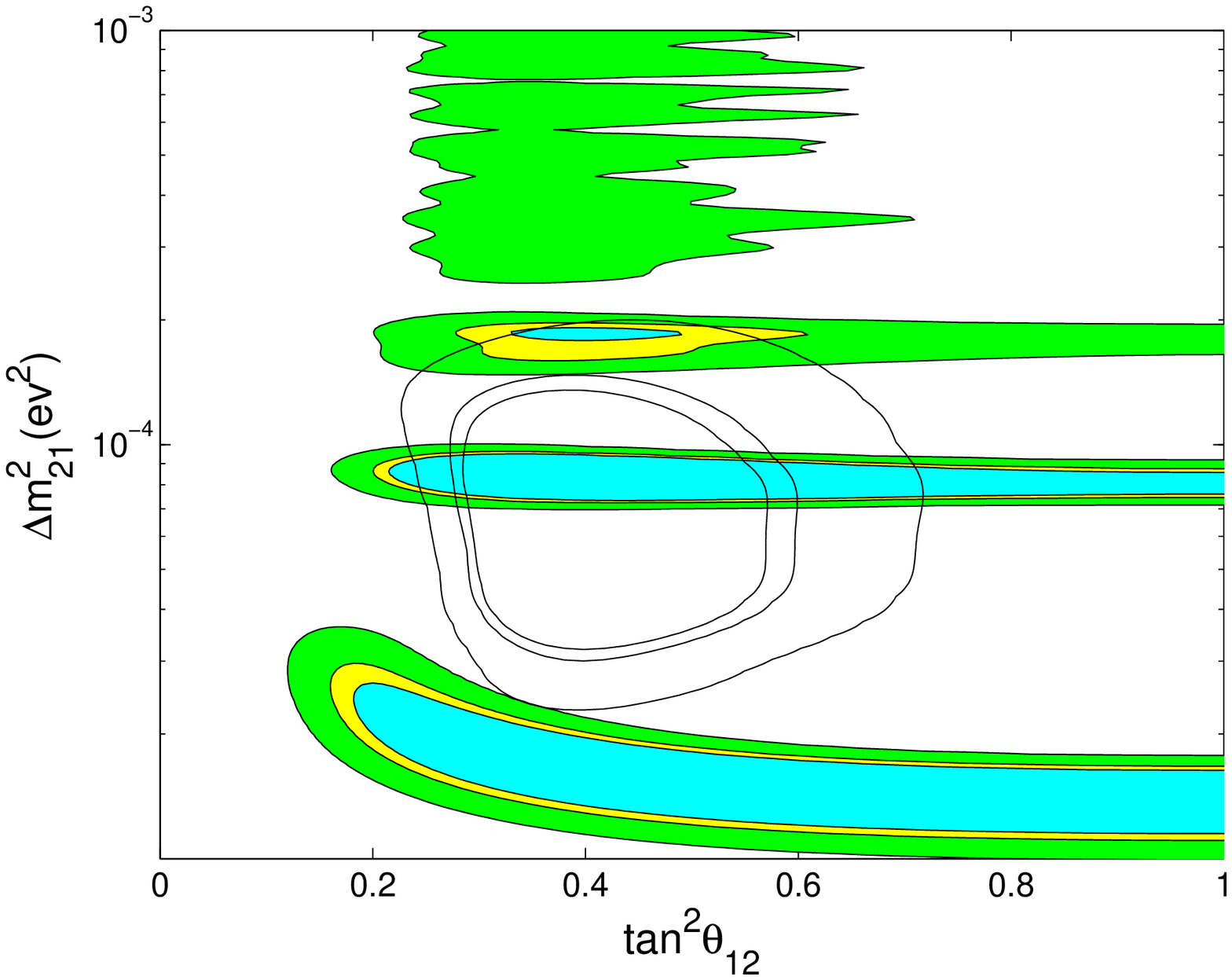,width=0.47\textwidth}}
&
\mbox{\epsfig{figure=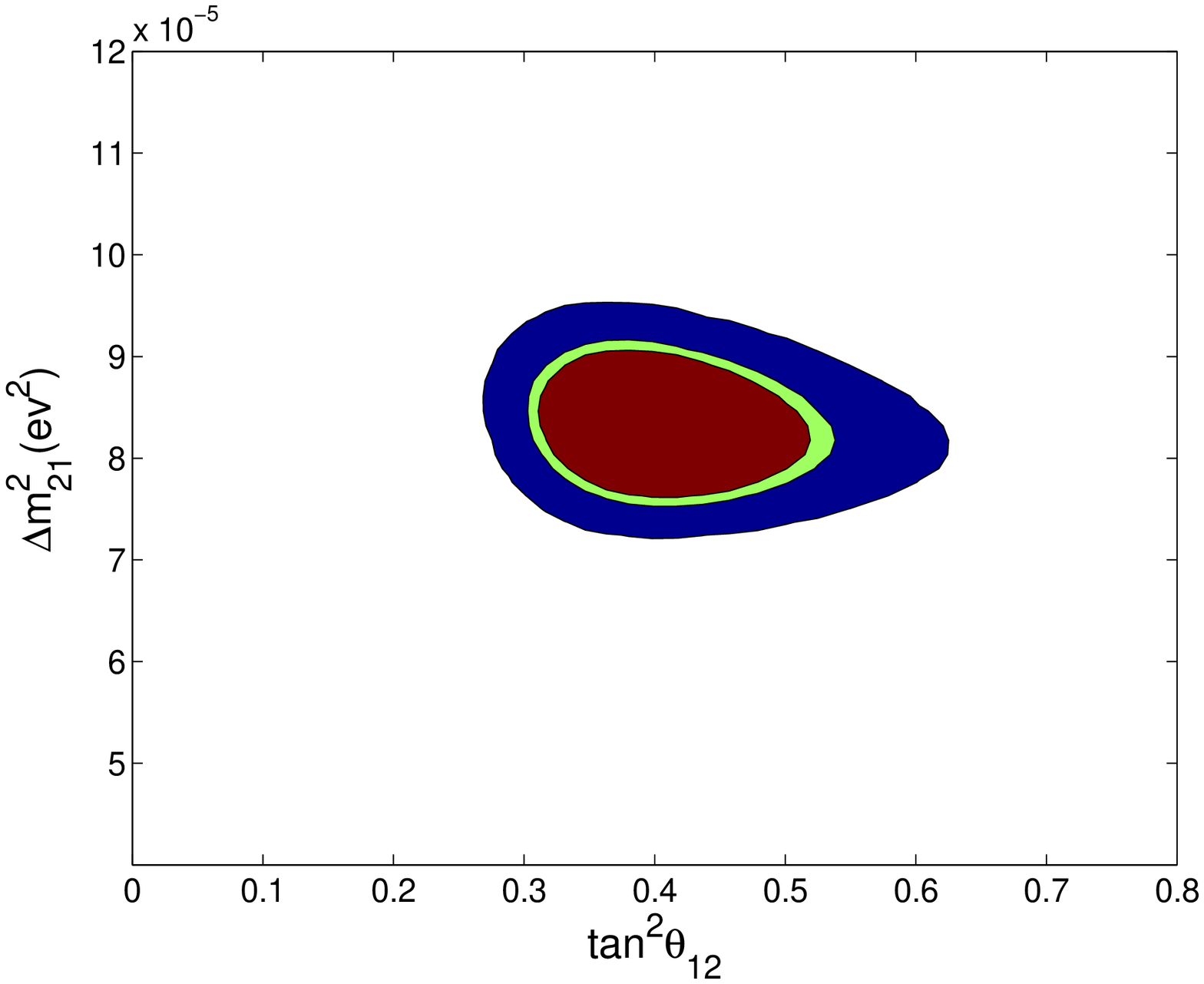,width=0.47\textwidth}}
\end{tabular}
\end{center}
\caption{
The 90\%, 95\%, 99.73\% Bayesian credible regions in the $\tan^2\!\theta_{12}$--$\Delta{m}^2_{21}$ plane.
Left: the regions
allowed by the recent KamLAND 766 ton-years data (shadowed regions),
together with the LMA region (empty contours) obtained from the analysis
the solar neutrino data.
Right: the region allowed by a combined analysis of
solar and KamLAND neutrino data.
}
\label{fig:bay}
\end{figure}

The prior distribution of the parameters,
$p(\tan^2\!\theta_{12},\Delta{m}^2_{21}|\mathcal{I})$,
is a subjective ingredient which quantifies the
knowledge on the values of the parameters which is independent from the
data to be analyzed.
In physics, it is generally considered desirable to
assume a prior distribution which carries as little information as possible
on the values of the parameters,
in order to obtain an unbiased result.
In this spirit,
we consider a flat prior distribution in the
$\tan^2\!\theta_{12}$--$\Delta{m}^2_{21}$ plane,
\begin{equation}
p(\tan^2\!\theta_{12},\Delta{m}^2_{21}|\mathcal{I}) = \mathrm{const}
\,.
\label{flatlin}
\end{equation}
In this case,
Eq.~(\ref{Bayes1}) becomes
\begin{equation}
p(\tan^2\!\theta_{12},\Delta{m}^2_{21}|\mathrm{D},\mathcal{I}) = \frac{
p(\mathrm{D}|\tan^2\!\theta_{12},\Delta{m}^2_{21},\mathcal{I}) \ }{
\displaystyle \int \mathrm{d}\!\tan^2\!\theta_{12} \,
\mathrm{d}\Delta{m}^2_{21} \,
p(\mathrm{D}|\tan^2\!\theta_{12},\Delta{m}^2_{21},\mathcal{I}) \ } \,.
\label{Bayes2}
\end{equation}
Using this expression, we obtained,
in the $\tan^2\!\theta_{12}$--$\Delta{m}^2_{21}$ plane,
the credible regions
with 90\%, 95\% and 99.73\% posterior probability shown in Fig.~\ref{fig:bay}.

In our Bayesian analysis, a credible region is
calculated by choosing the smallest region over which the
integral of the posterior probability distribution function is the
given probability level.
For example,
a 90\% Bayesian credible region is the smallest region for which
the posterior probability of the parameters is 90\%.

\begin{figure}[t!]
\begin{center}
\begin{tabular}{ccc}
\mbox{\epsfig{figure=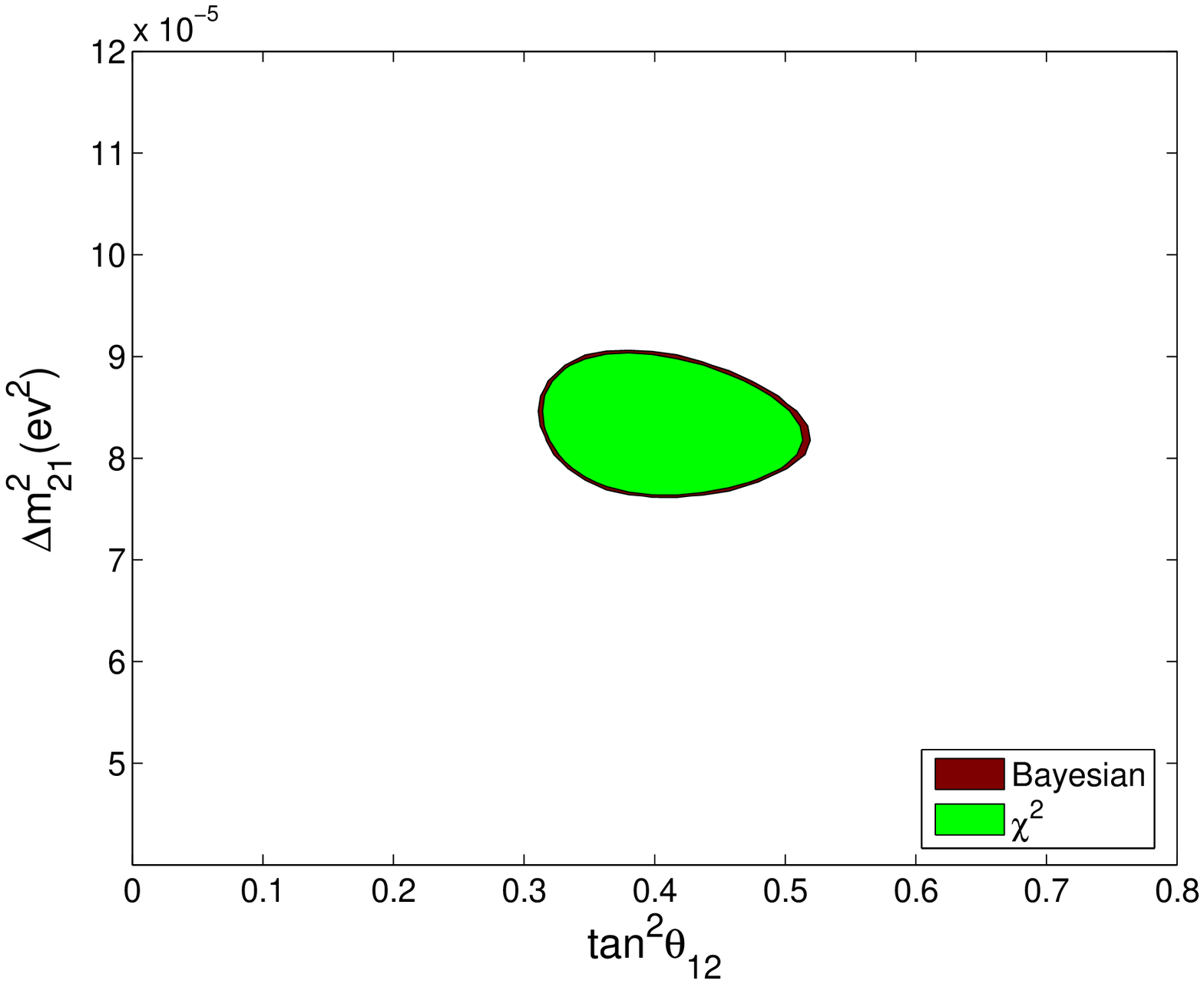,width=0.30\textwidth}}
&
\mbox{\epsfig{figure=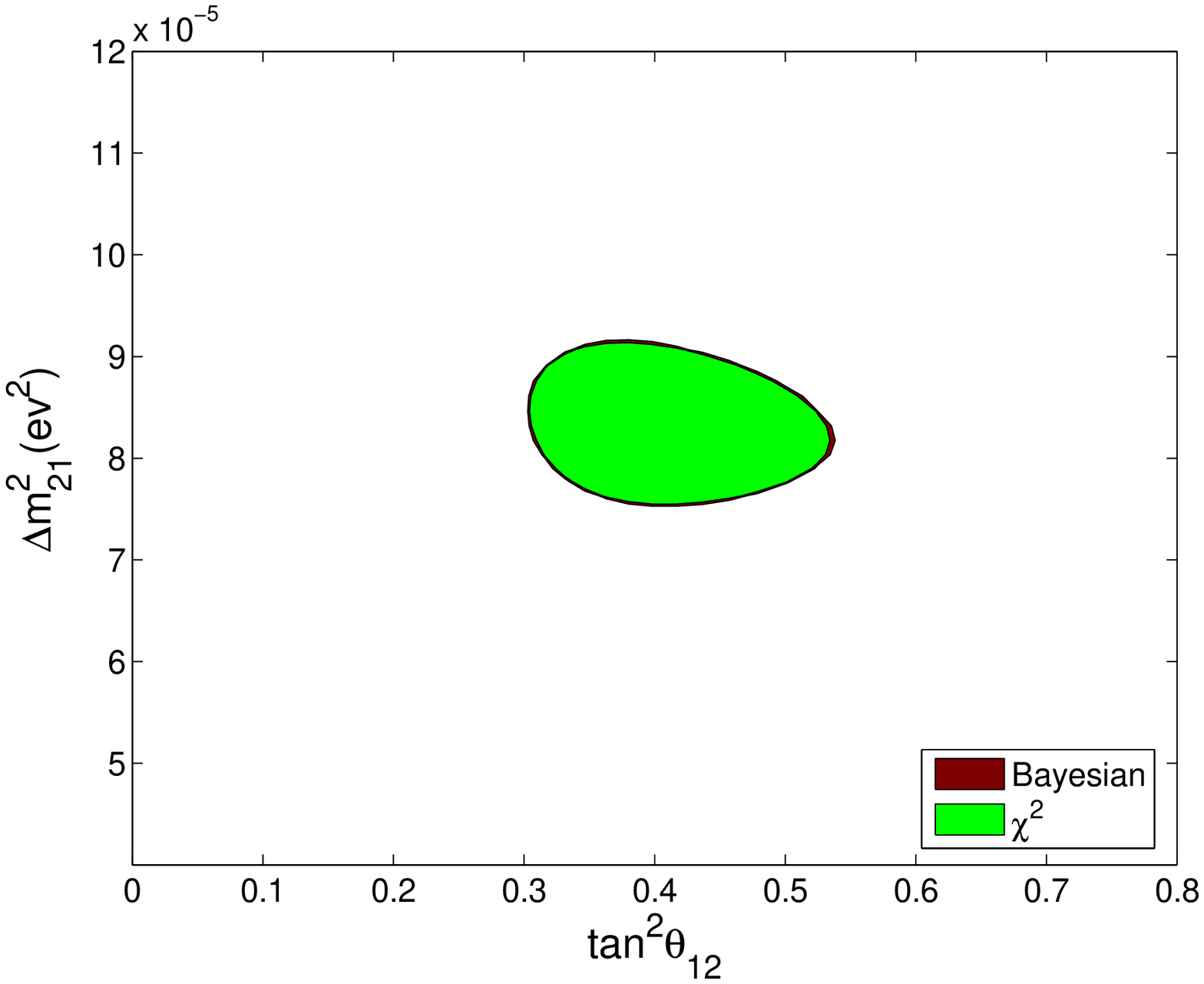,width=0.30\textwidth}}
&
\mbox{\epsfig{figure=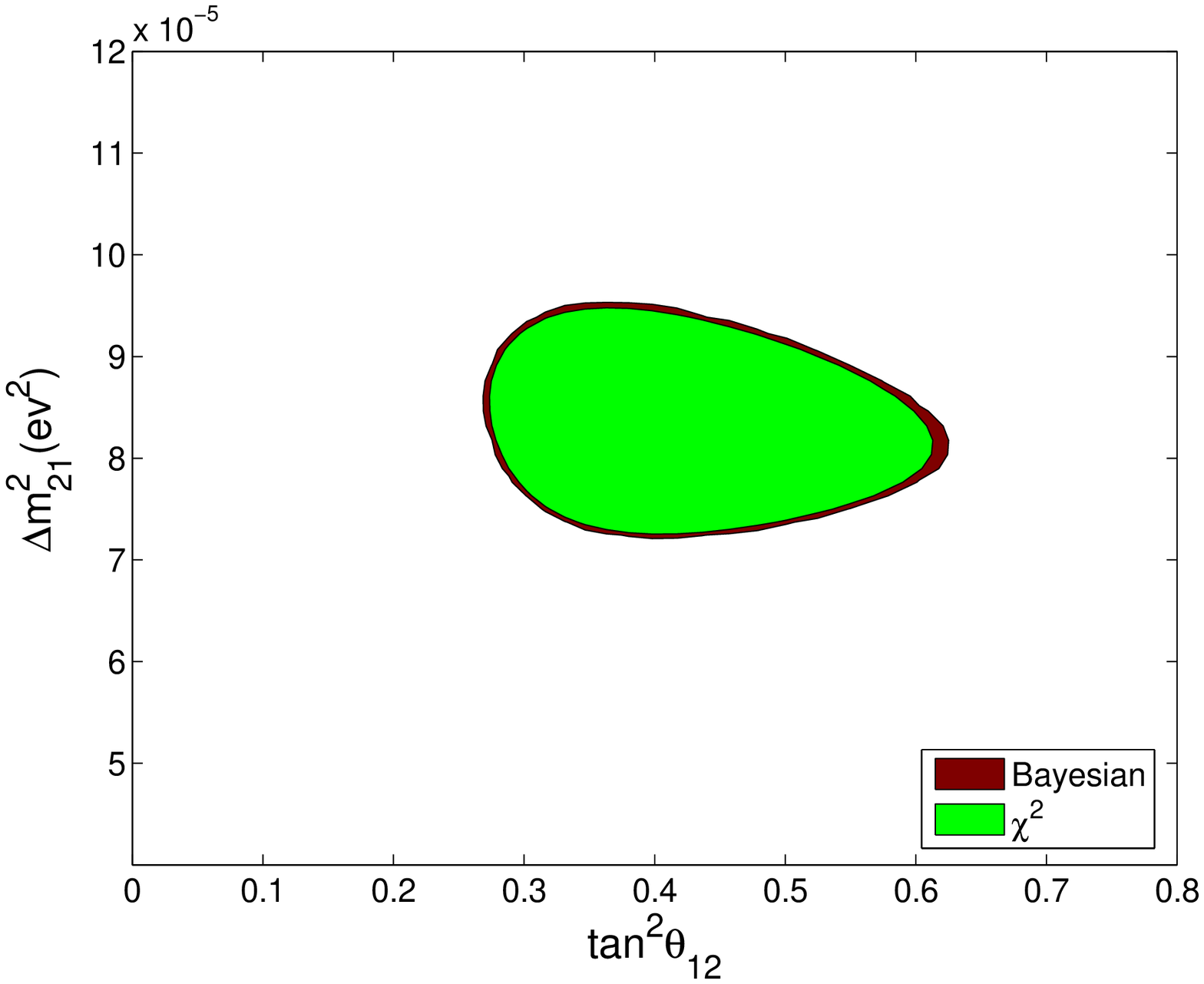,width=0.30\textwidth}}
\end{tabular}
\end{center}
\caption{Comparisons of
the Bayesian and least-squares methods.
Going from left to right,
the three figures correspond to
90\%, 99\% and 99.73\% probability (Bayesian) or
confidence level (least-squares). } \label{fig:compare}
\end{figure}

Let us emphasize that the statistical meanings of Bayesian and
Frequentist regions are different (see
Refs.~\cite{Jeffreys-book-39,Eadie-71,Kendall-2A,Loredo-90,Loredo-92,Jaynes-book,D'Agostini-book,hep-ph/0007155,bay2}).
Nevertheless, it is interesting to compare the allowed regions
obtained with the Bayesian and least-squares methods, especially
in view of the fact that many scientists (and most human beings)
do not know the meaning of Frequentist allowed regions and tend to
give them a probability content, as if they were Bayesian credible
regions. Fig.~\ref{fig:compare} shows a comparison of the Bayesian
credible regions with 90\%, 95\% and 99.73\% probability with the
Frequentist allowed regions at 90\%, 95\% and 99.73\% confidence
level obtained with a least-squares analysis. One can see that the
Bayesian credible regions are similar but a little larger than the
corresponding allowed regions obtained with the least-squares
analysis.

Differences between Bayesian credible regions and the corresponding
Frequentist allowed regions can be due
to a poor statistical quality of the data
or to a wrong theoretical model which leads to a bad fit of the data.
The close similarity of the Bayesian credible regions
and the corresponding
Frequentist allowed regions
that we have obtained is a signal of the
very good statistical quality of the data
and a confirmation of the LMA solution of the solar neutrino problem.
Note,
however,
that the similarity of the Bayesian and Frequentist
results depends on the assumption of a flat prior in the Bayesian analysis.
In section~\ref{sec:work}, we
will change the prior probability distribution function in order to further test
the stability of the LMA solution.

It is also useful to calculate the marginal
posterior probability
distributions of $\tan^2\!\theta_{12}$ and $\Delta{m}^2_{21}$,
which are given by
\begin{eqnarray}
p(\tan^2\!\theta_{12}|\mathrm{D},\mathcal{I}) = \int
\mathrm{d}\Delta{m}^2_{21} \,
p(\tan^2\!\theta_{12},\Delta{m}^2_{21}|\mathrm{D},\mathcal{I}) \,,
\label{margial1}
\\
p(\Delta{m}^2_{21}|\mathrm{D},\mathcal{I}) = \int
\mathrm{d}\!\tan^2\!\theta_{12} \,
p(\tan^2\!\theta_{12},\Delta{m}^2_{21}|\mathrm{D},\mathcal{I}) \,.
\label{marginal2}
\end{eqnarray}
These distributions give information
on each of the two parameters independently from the value of the other.

\begin{figure}[t!]
\begin{center}
\begin{tabular}{cc}
\mbox{\epsfig{figure=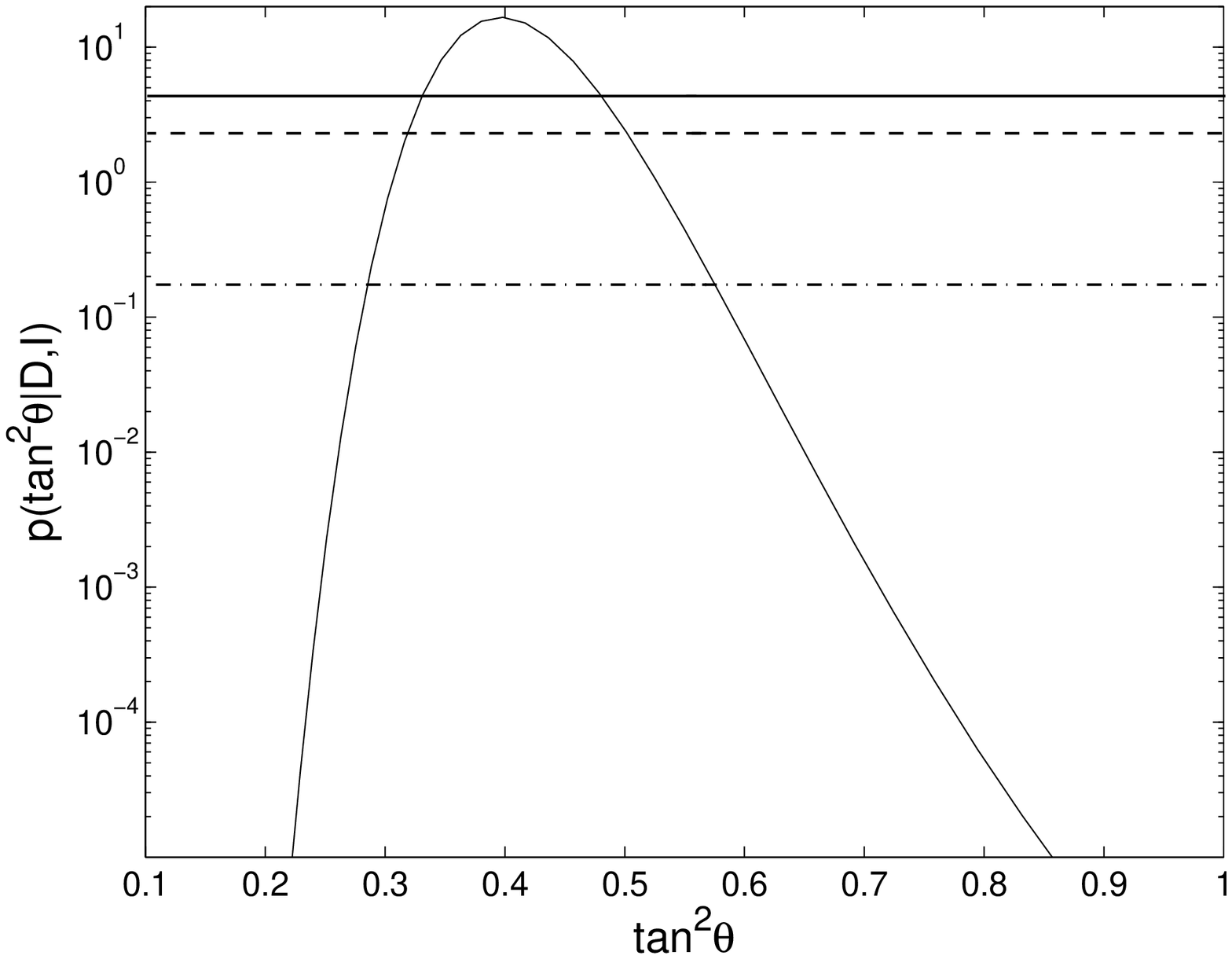,width=0.47\textwidth}}
&
\mbox{\epsfig{figure=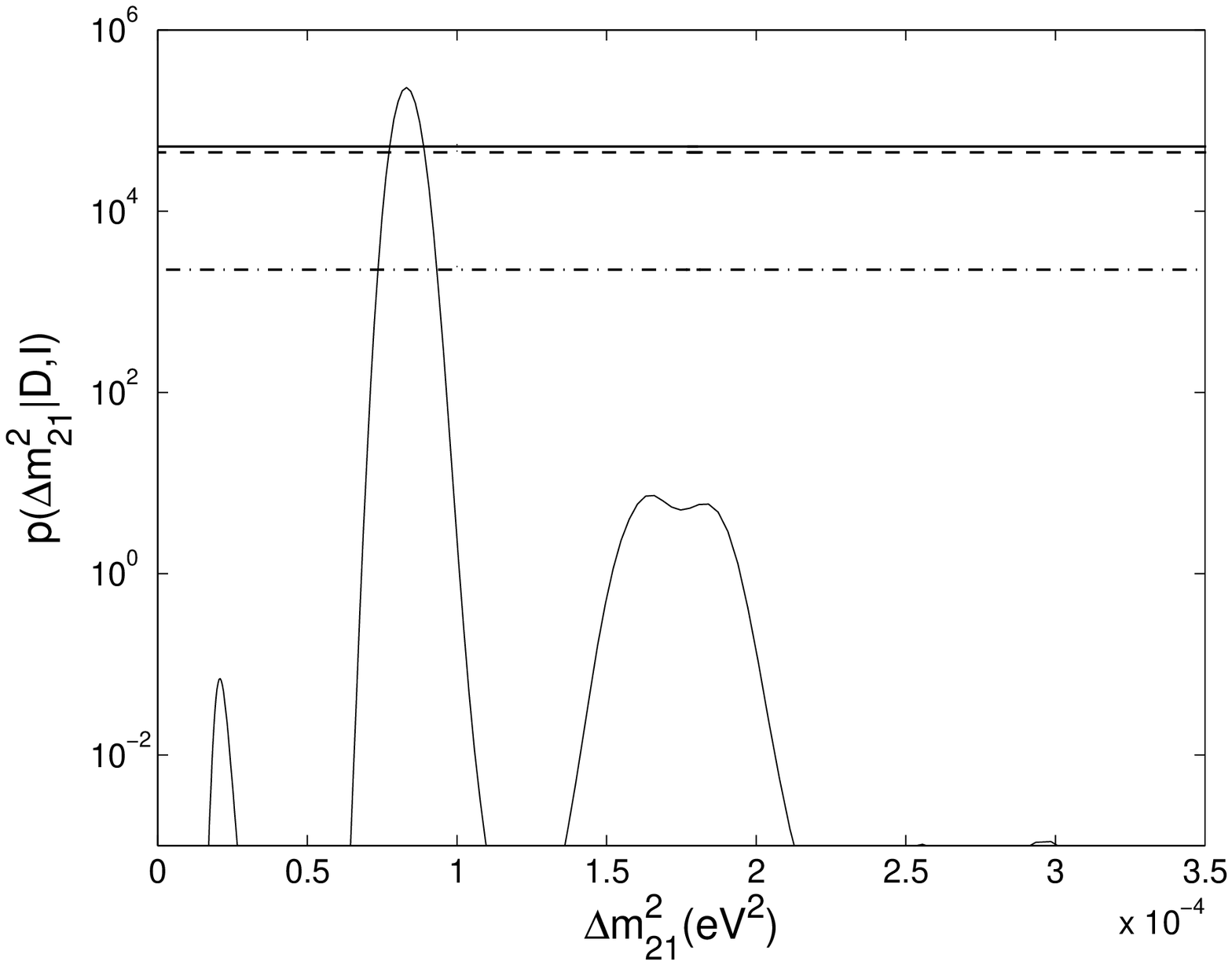,width=0.47\textwidth}}
\end{tabular}
\end{center}
\caption{The marginal
Bayesian distributions: left figure is a separated posterior
probability distribution for $\tan^2\!\theta_{12}$; right figure is the
separated posterior probability distributions for $\Delta{m}^2_{21}$.
Three horizontal lines in each figure show three integrated
probabilities: 90\%, 95\% and 99.73\%(from up to down
respectively).
}
\label{fig:marg}
\end{figure}

The marginal posterior probability distributions
corresponding to the credible regions in Fig.~\ref{fig:bay}
are shown in Fig.~\ref{fig:marg}.

The marginal posterior probability distribution for $\tan^2\!\theta_{12}$,
which is mainly determined by solar data,
has only one peak with an approximate skewed-Gaussian shape.
The resulting value of $\tan^2\!\theta_{12}$ is
\begin{equation}
\tan^2\!\theta_{12} = 0.40 {}^{+0.04}_{-0.04} \quad \text{(68\%
probability range)} \,. \label{thetabf}
\end{equation}
The allowed intervals with 90\%, 95\% and 99.73\% probability,
each of which is given by the intersections of the distribution
function with the corresponding horizontal line
in the left panel of Fig.~\ref{fig:marg},
are
\begin{equation}
\tan^2\!\theta_{12} = [ 0.33 \,,\, 0.48 ] \, (90\%) \,, \quad [
0.32 \,,\, 0.50 ] \, (95\%) \,, \quad [ 0.29 \,,\, 0.57 ] \,
(99.73\%) \,. \label{thetaint}
\end{equation}

The marginal posterior probability distribution for $\Delta{m}^2_{21}$
has one main peak and two small peaks on the sides.
These two minor peaks correspond to the regions in the
left panel of Fig.~\ref{fig:bay} which are allowed by the KamLAND data
and have a partial overlap with the solar allowed region.
However,
from the right panel in Fig.~\ref{fig:marg},
one can see that they are strongly disfavored with respect to the main peak,
which gives
\begin{equation}
\Delta{m}^2_{21} = 8.32{}^{+0.29}_{-0.30}\times10^{-5} \,
\text{eV}^2 \quad \text{(68\% probability range)} \,.
\label{dm2bf}
\end{equation}
The allowed intervals with 90\%, 95\% and 99.73\% probability
are
\begin{equation}
\Delta{m}^2_{21} = [ 7.76 \,,\, 8.89 ] \, (90\%) \,, \quad [ 7.73
\,,\, 8.91 ] \, (95\%) \,, \quad [ 7.37 \,,\, 9.32 ] \, (99.73\%)
 \quad \times10^{-5}\text{eV}^2 \,. \label{dm2int}
\end{equation}

Let us emphasize that the remarkable precision of the determination of
$\tan^2\!\theta_{12}$ and $\Delta{m}^2_{21}$
in Eqs.~(\ref{thetabf})--(\ref{dm2int})
represents an impressive success of the solar and KamLAND neutrino experiments.


\section{Stability of the LMA region}
\label{sec:work}

The Bayes' Theorem in Eq.~(\ref{Bayes1}) requires a
prior probability distribution function.
The flat prior that we have adopted in the previous section
is often considered as the best non-informative choice.
In this section, we explore the implications for the
Bayesian credible regions of different
choices for the prior distribution, which may quantify
some prior belief or knowledge.

From theoretical considerations, it is know that flavor
transitions of solar neutrinos can occur over several orders of
magnitude of $\tan^2\!\theta_{12}$ and $\Delta{m}^2_{21}$, through
vacuum oscillations for $\Delta{m}^2_{21} \lesssim 10^{-8} \,
\text{eV}^2$ and large mixing angles or resonant MSW transitions
for $ 10^{-8} \, \text{eV}^2 \lesssim \Delta{m}^2_{21} \lesssim
10^{-3} \, \text{eV}^2 $ and $ 10^{-4} \lesssim
\tan^2\!\theta_{12} \lesssim 10 $. Hence, one may think that a
flat prior distribution in the
$\log(\tan^2\!\theta_{12})$--$\log(\Delta{m}^2_{21})$ plane,
\begin{equation}
p(\log(\tan^2\!\theta_{12}),\log(\Delta{m}^2_{21})|\mathcal{I}) = \mathrm{const}
\,,
\label{flatlog1}
\end{equation}
is more appropriate than the flat prior distribution in the
$\tan^2\!\theta_{12}$--$\Delta{m}^2_{21}$ plane in
Eq.~(\ref{flatlin}). In fact, this prior was adopted in
Refs.~\cite{bay2,Creminelli:2001ij,model}. In the
$\tan^2\!\theta_{12}$--$\Delta{m}^2_{21}$ plane, it roughly
corresponds to the prior
\begin{equation}
p(\tan^2\!\theta_{12},\Delta{m}^2_{21}|\mathcal{I}) \propto \frac{
1 }{ \tan^2\!\theta_{12} \, \Delta{m}^2_{21} } \,.
\label{flatlog2}
\end{equation}
Slight differences between Eq.~\ref{flatlog1} and
Eq.~\ref{flatlog2} is that in the
$\tan^2\!\theta_{12}$--$\Delta{m}^2_{21}$ plane the credible
region by Eq.~\ref{flatlog1} is no more iso contours and the
probability at the center is shifted from the biggest value, and
vise versa.

\begin{figure}[t!]
\begin{center}
\begin{tabular}{cc}
\mbox{\epsfig{figure=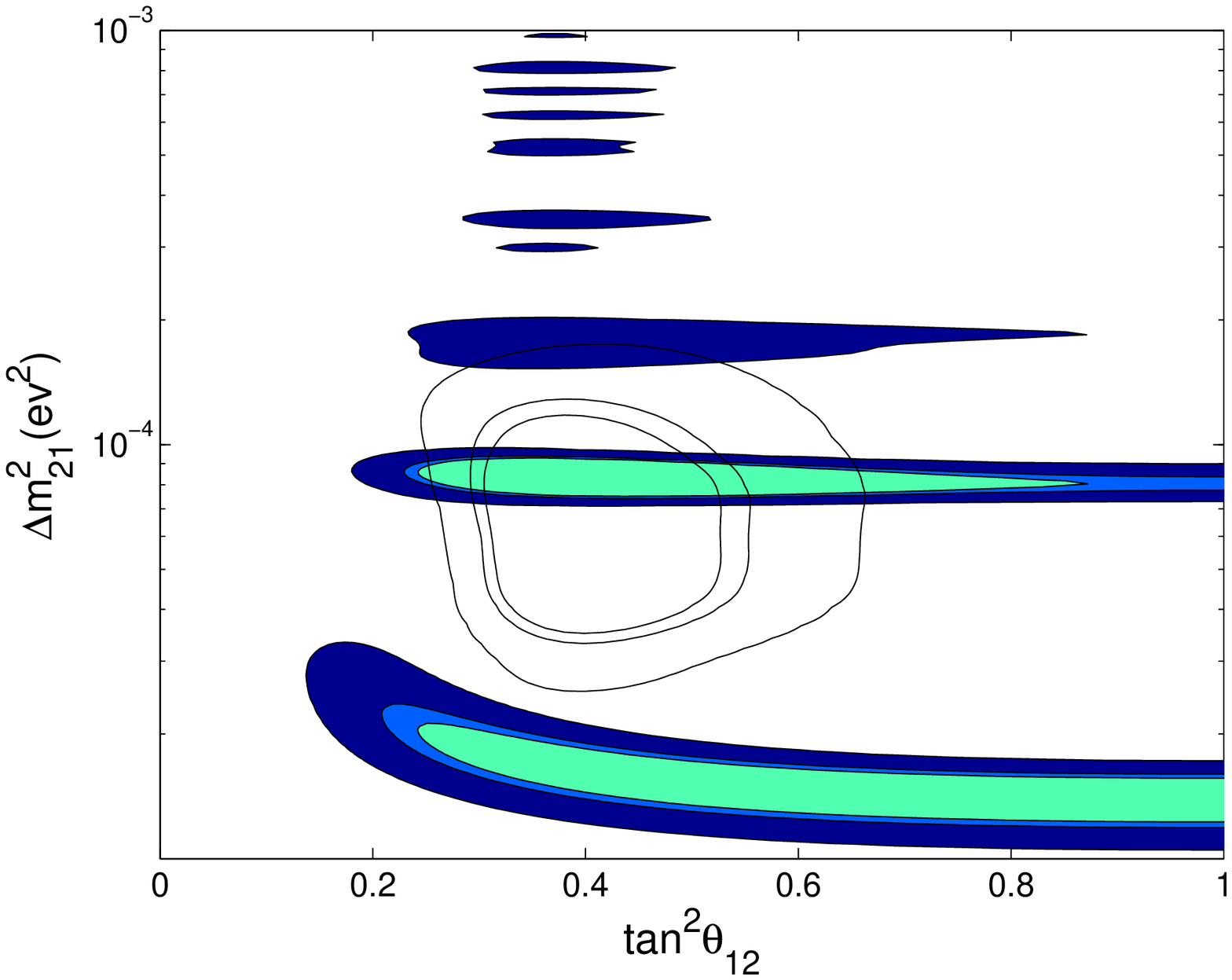,width=0.47\textwidth}}
&
\mbox{\epsfig{figure=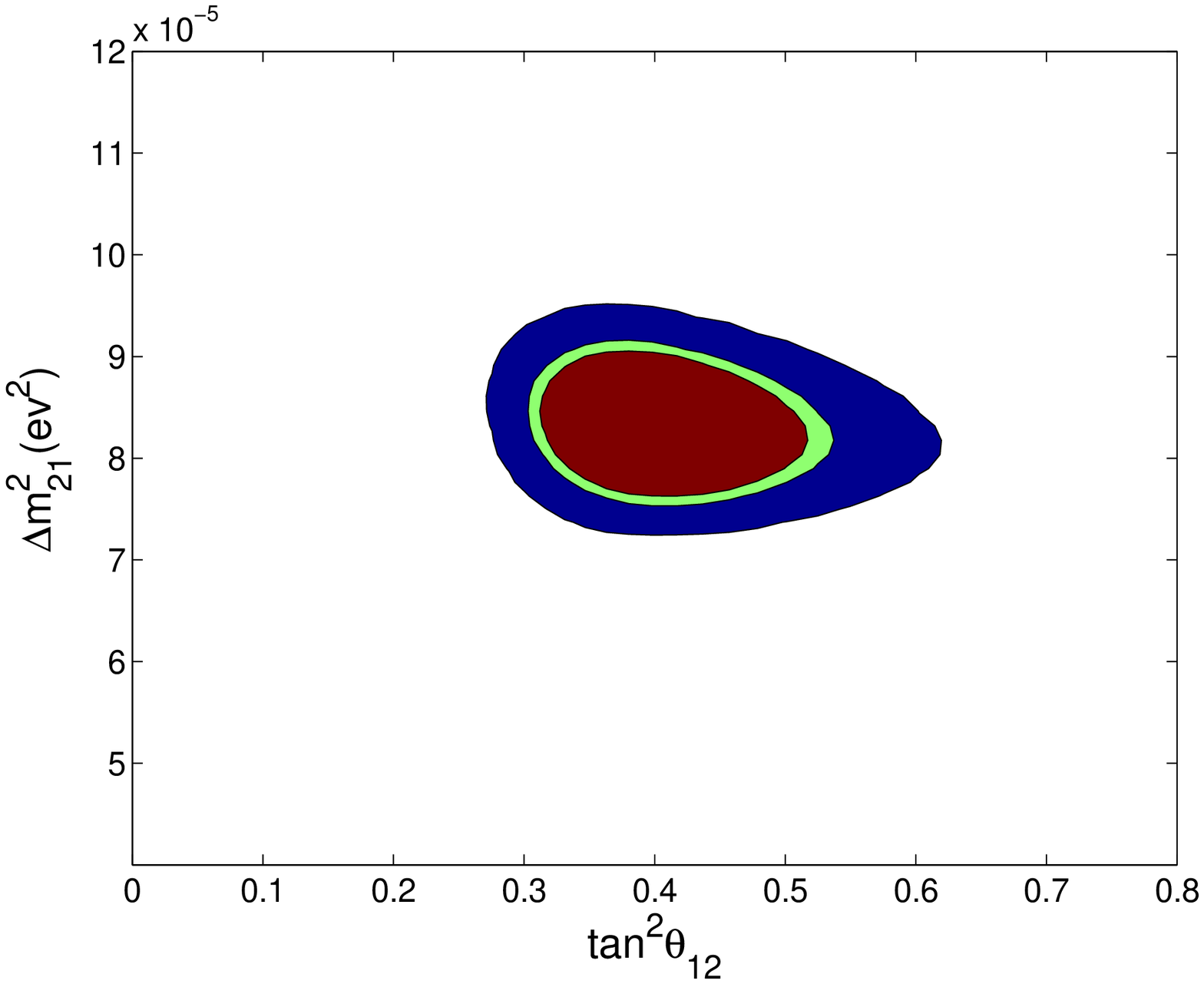,width=0.47\textwidth}}
\end{tabular}
\end{center}
\caption{ The 90\%, 95\%, 99.73\% Bayesian credible regions in the
$\log(\tan^2\!\theta_{12})$--$\log(\Delta{m}^2_{21})$ plane
obtained with the prior in Eq.~(\ref{flatlog1}). Left: the regions
allowed by the KamLAND 766 ton-years data (shadowed regions),
together with the LMA region (empty contours) obtained from the
analysis the solar neutrino data. Right: the region allowed by a
combined analysis of solar and KamLAND neutrino data. }
\label{fig:baylog}
\end{figure}

Fig.~\ref{fig:baylog} shows the credible regions with 90\%, 95\%
and 99.73\% posterior probability obtained with the prior in
Eq.~(\ref{flatlog1}). One can see that they are rather similar to
the ones in Fig.~\ref{fig:bay}, which have been obtained with the
prior in Eq.~(\ref{flatlin}), except for a slight preference of
small values of the parameters due to the prior. However, a
comparison of the right panels in Figs.~\ref{fig:bay} and
\ref{fig:baylog} leads to the conclusion that the data are so good
that the priors in Eqs.~(\ref{flatlin}) and (\ref{flatlog2})
produce almost the same result for the credible regions. Hence, in
practice the choice between the priors in Eqs.~(\ref{flatlin}) and
(\ref{flatlog2}) is irrelevant.

We consider now a prior
which is a flat distribution with respect to $\Delta{m}^2_{21}$ and a normal distribution
for the parameter $\tan^2\!\theta_{12}$:
\begin{equation}
p(\tan^2\!\theta_{12},\Delta{m}^2_{21}|\mathcal{I}) = \frac{ 1 }{ \sqrt{2\pi}\sigma}
\,
e^{-(\tan^2\theta_{12}-\mu)^2/2\sigma^2}
\,.
\label{B1}
\end{equation}
This case corresponds to a prior belief in $\tan^2\!\theta_{12} \simeq \mu$,
with an uncertainty $\sigma$.
The Bayes' Theorem in Eq.~(\ref{Bayes1}) gives the posterior probability distribution
\begin{equation}
p(\tan^2\!\theta_{12},\Delta{m}^2_{21}|\mathrm{D},\mathcal{I}) = \frac{ \displaystyle
p(\mathrm{D}|\tan^2\!\theta_{12},\Delta{m}^2_{21},\mathcal{I})
\,
e^{-(\tan^2\theta_{12}-\mu)^2/2\sigma^2}
}{ \displaystyle
\int \mathrm{d}\!\tan^2\!\theta_{12} \,
\mathrm{d}\Delta{m}^2_{21} \,
p(\mathrm{D}|\tan^2\!\theta_{12},\Delta{m}^2_{21},\mathcal{I}) \,
e^{-(\tan^2\theta_{12}-\mu)^2/2\sigma^2}
}
\label{B2}
\,.
\end{equation}

\begin{figure}[t!]
\begin{center}
\begin{tabular}{ccc}
\mbox{\epsfig{figure=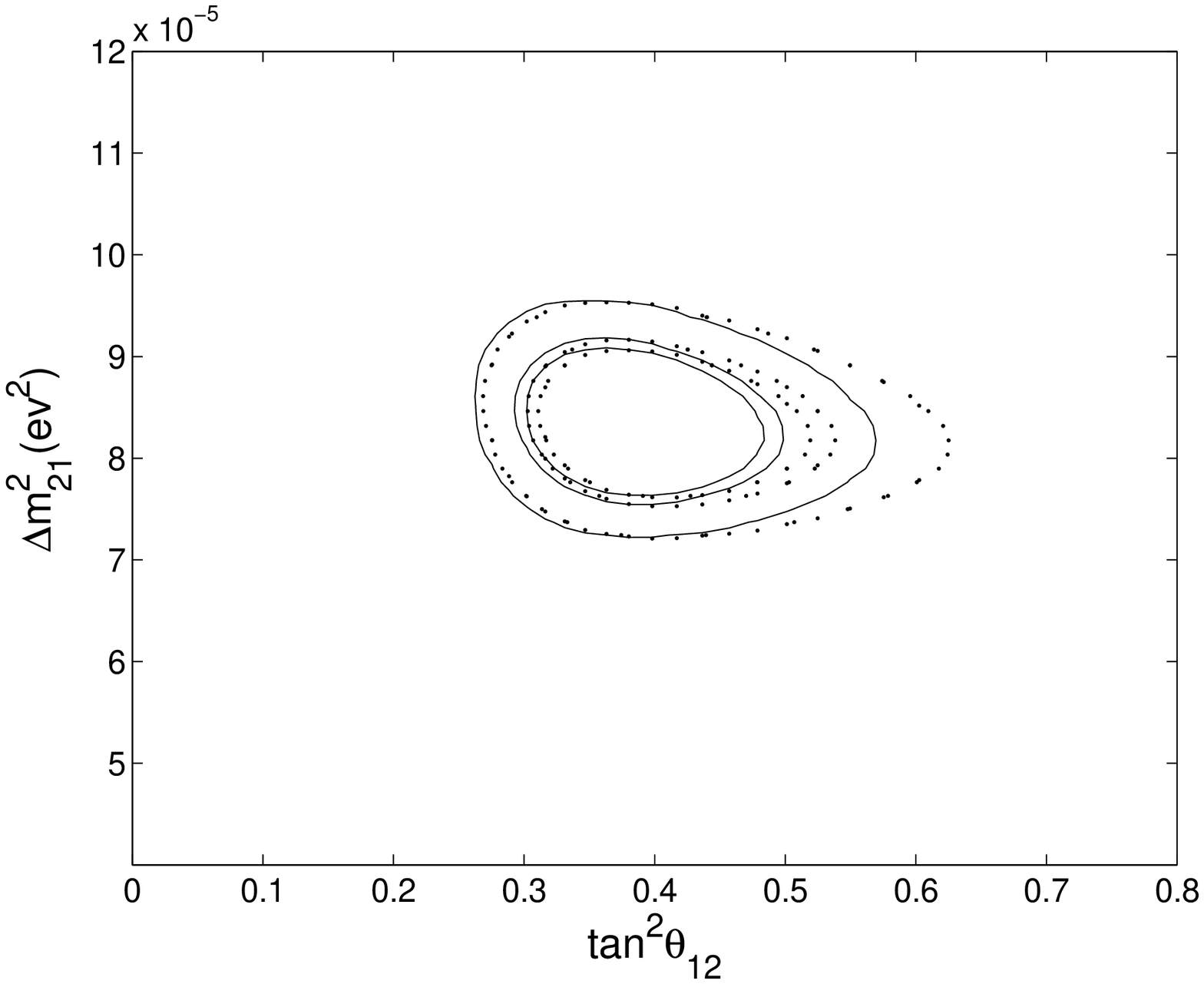,width=0.30\textwidth}}
&
\mbox{\epsfig{figure=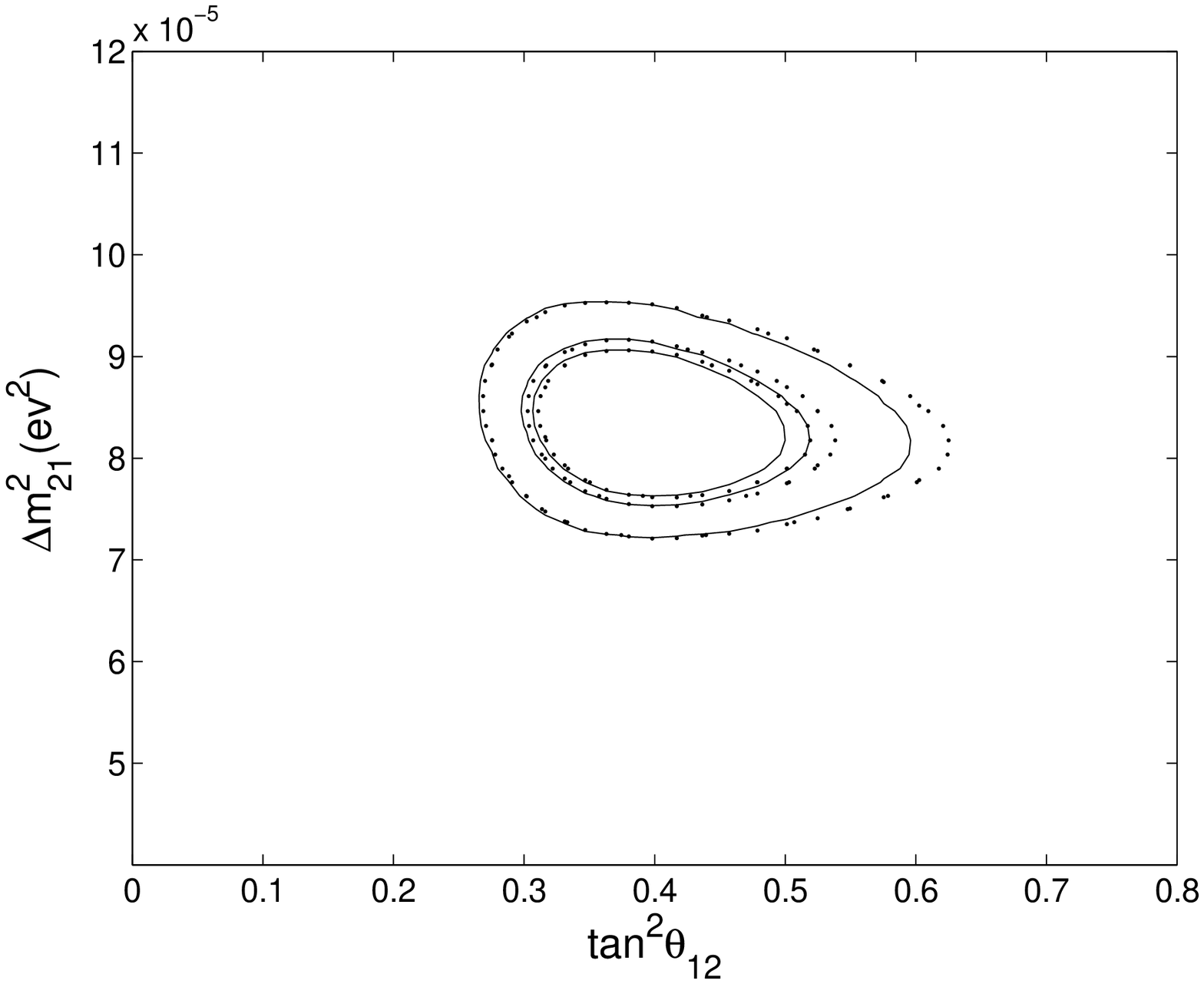,width=0.30\textwidth}}
&
\mbox{\epsfig{figure=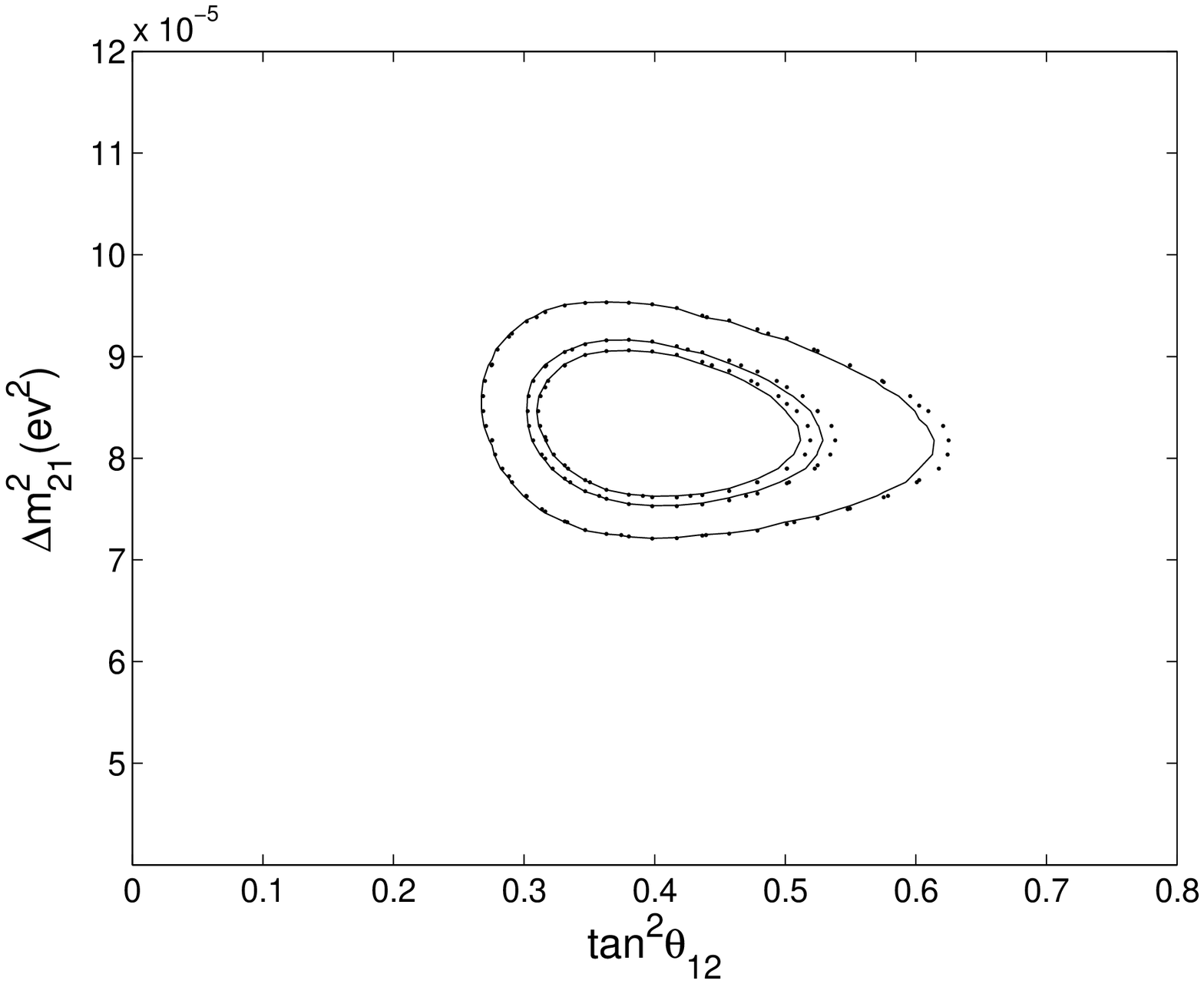,width=0.30\textwidth}}
\end{tabular}
\end{center}
\caption{
The 90\%, 95\%, and 99.73\% credible regions corresponding to the posterior
probability distribution
function in Eq.~(\ref{B2}), with $\mu=0$.
The left, middle and right figures have been obtained, respectively, with
$\sigma=0.2$, $0.3$, and $0.5$.
The dotted lines correspond to a flat prior distribution
(same as the right figure in Fig.~\ref{fig:bay}).
}
\label{fig:gaus0}
\end{figure}

Fig.~\ref{fig:gaus0} shows the 90\%, 95\%, and 99.73\% credible
regions for $\mu = 0$ and $\sigma=0.2 , 0.3 , 0.5$. The value $\mu
= 0$ corresponds to a prior belief that the mixing angle should be
small. One can see from the three panels in Fig.~\ref{fig:gaus0}
that a small value of $\sigma$ has the effect to exclude the
large-$\tan^2\theta_{12}$ part of the credible regions obtained
with a flat prior in section~\ref{sec:bayesian}. On the other
hand, the low-$\tan^2\theta_{12}$ part of the credible regions is
not affected by the change of prior. This is due to the KamLAND
data, which exclude small values of $\tan^2\theta_{12}$, as shown
in the left panel of Fig.~\ref{fig:bay}.

\begin{figure}[t!]
\begin{center}
\begin{tabular}{ccc}
\mbox{\epsfig{figure=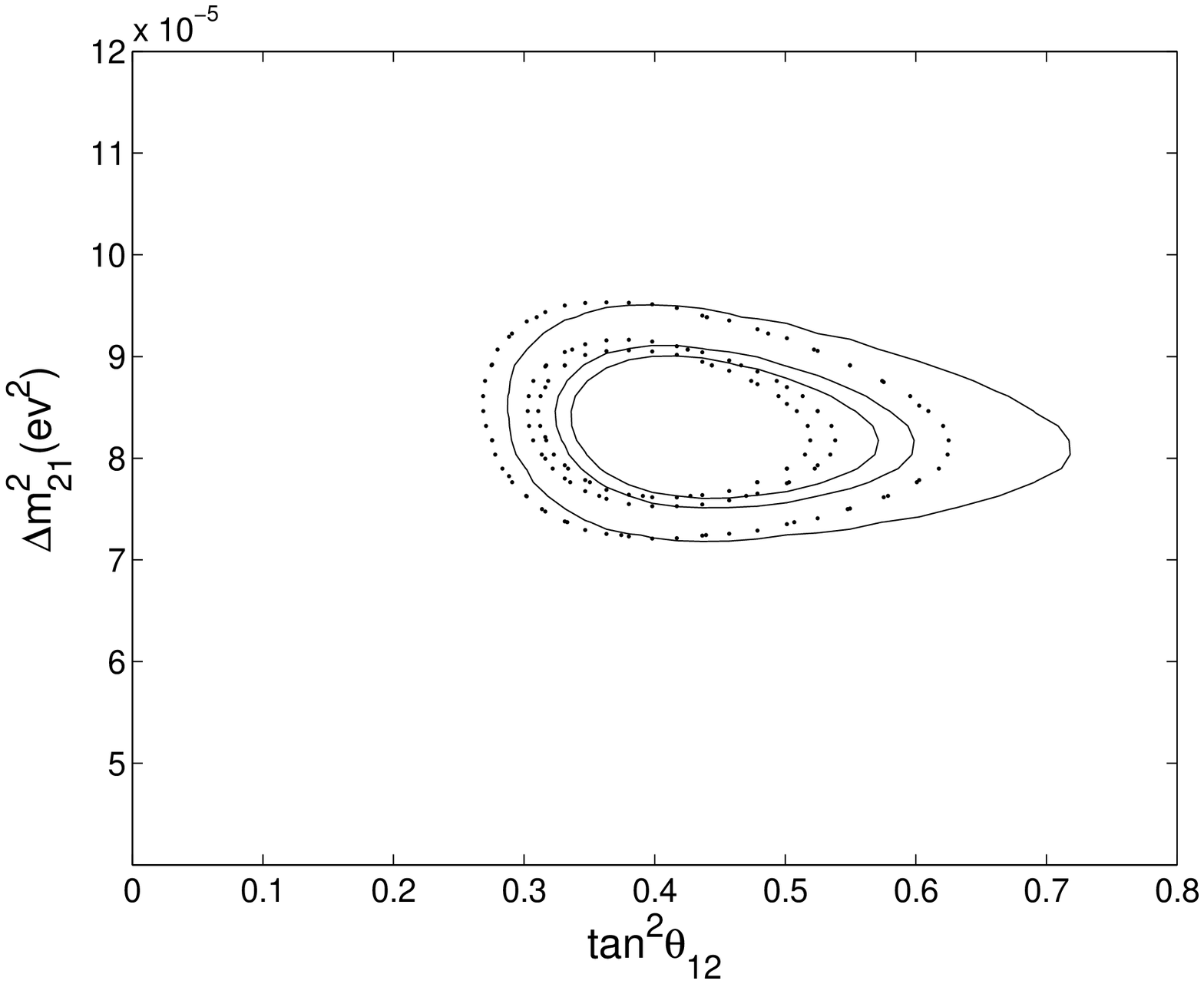,width=0.30\textwidth}}
&
\mbox{\epsfig{figure=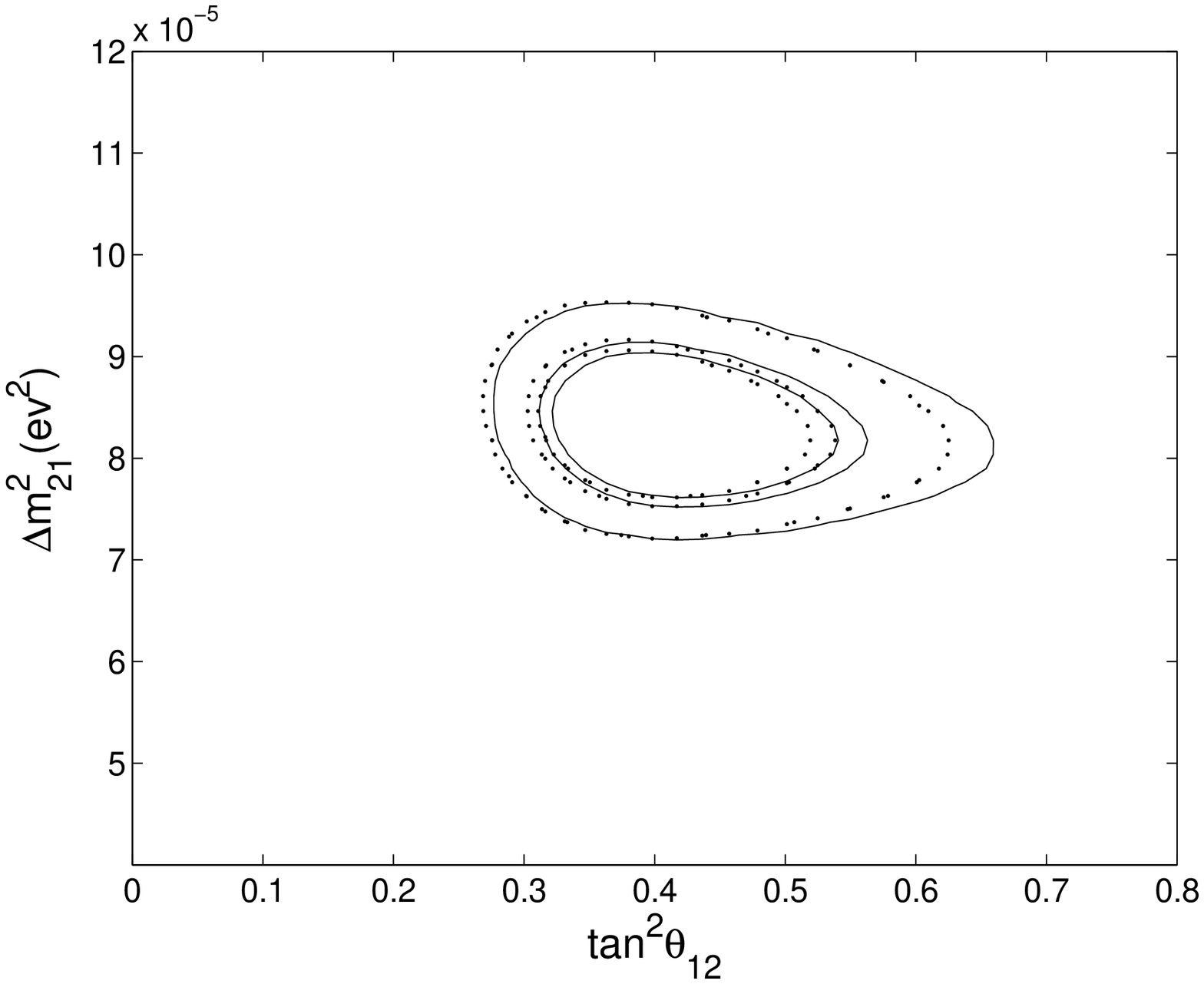,width=0.30\textwidth}}
&
\mbox{\epsfig{figure=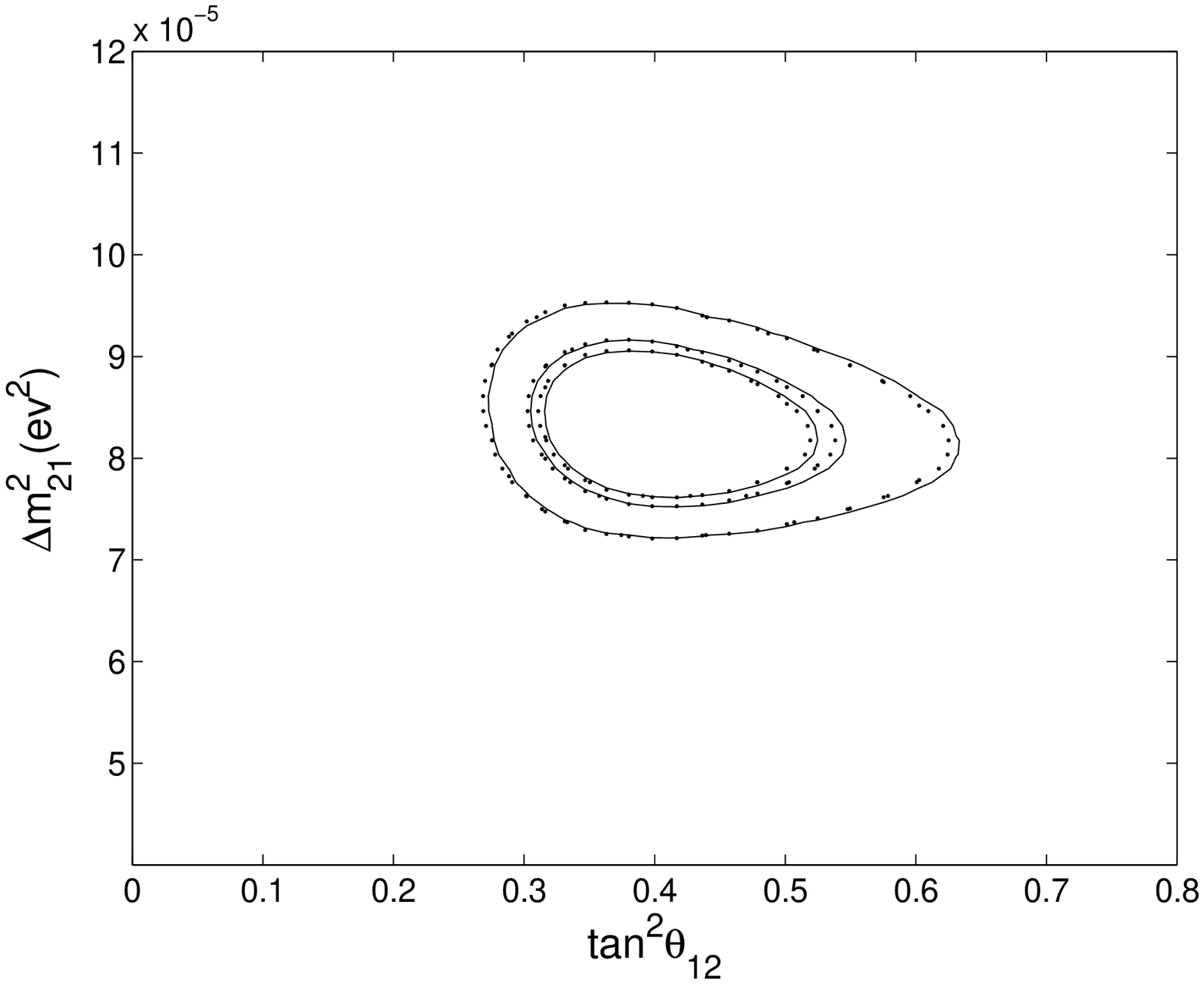,width=0.30\textwidth}}
\end{tabular}
\end{center}
\caption{
The 90\%, 95\%, and 99.73\% credible regions corresponding to the posterior
probability distribution
function in Eq.~(\ref{B2}), with $\mu=1$.
The left, middle and right figures have been obtained, respectively, with
$\sigma=0.2$, $0.3$, and $0.5$.
The dotted lines correspond to a flat prior distribution
(same as the right figure in Fig.~\ref{fig:bay}).
}
\label{fig:gaus1}
\end{figure}

Figs.~\ref{fig:gaus1} shows the 90\%, 95\%, and 99.73\% credible
regions for $\mu = 1$ and $\sigma=0.2 , 0.3 , 0.5$. The value $\mu
= 1$ corresponds to a prior belief in favor of a large mixing
angle, close to maximal. The leftmost panel in
Fig.~\ref{fig:gaus1} shows that, with respect to the case of a
flat prior considered in section~\ref{sec:bayesian}, a small value
of $\sigma$ leads to an enlargement of the credible regions
towards large values of $\tan^2\theta_{12}$, which are allowed by
the KamLAND data, as shown in the left panel of
Fig.~\ref{fig:bay}. The low-$\tan^2\theta_{12}$ part of the
credible regions is affected only mildly by the change of prior.

In conclusion,
in this section we have shown that the LMA solution is stable
when reasonable priors are chosen
in place of the flat one considered in section~\ref{sec:bayesian},
with a possible shrink or enlargement of the
large-$\tan^2\theta_{12}$ part of the credible regions
depending on a prior belief in favor of
a small or large mixing angle.

\section{Conclusions}
\label{sec:conclusions}

In this paper, we
presented the results of a
Bayesian
analysis of the solar and KamLAND neutrino data.
We showed that the Bayesian
analysis with a flat prior distribution in the
$\tan^2\!\theta_{12}$--$\Delta{m}^2_{21}$ plane
leads to an allowed LMA region for $\nu_e\to\nu_{\mu,\tau}$ transitions
which practically coincides
with the one obtained with a standard least-squares ($\chi^2$)
analysis.
We investigated the stability of the LMA allowed region
for other reasonable choices of the prior.
We have shown that the LMA solution is stable
against reasonable variations of the prior,
with a possible shrink or enlargement of the
large-$\tan^2\theta_{12}$ part of the allowed LMA region
if one has a prior belief, respectively, in favor of
a small or large mixing angle.

\section{Acknowledgments}

The authors from USTC would like to thank M.J. Luo and Y.F. Li for
useful discussions. This work is supported in part by the National
Natural Science Foundation of China under grant number 90203002.

\end{document}